# Domain wall propagation by spin-orbit torques in in-plane magnetized systems

R. Kohno[1], J. Sampaio[1,*], S. Rohart[1], and A. Thiaville[1]

[1] *Université Paris-Saclay, CNRS, Laboratoire de Physique des Solides, Orsay, France*
* joao.sampaio@u-psud.fr

The effect of damping-like spin-orbit torque (DL SOT) on magnetic domain walls (DWs) in in-plane magnetised soft tracks is studied analytically and with micromagnetic simulations. We find that DL SOT drives vortex DWs, whereas transverse DWs, the other typical DW structure in soft tracks, propagate only if the Dzyaloshinskii-Moriya interaction (DMI) is present. The SOT drive can add to, and be more efficient than, spin-transfer torque (STT), and so may greatly benefit applications that require in-plane DWs. Our analysis based on the Thiele equation shows that the driving force arises from a cycloidal distortion of the DW structure caused by geometrical confinement or DMI. This distortion is higher, and the SOT more efficient, in narrower, thinner tracks. These results show that the effects of SOT cannot be understood by simply considering the effective field at the center of the structure, an ill-founded but often-used estimation. We also show that the relative magnitude of STT and DL SOT can be determined by comparing the motion of different vortex DW structures in the same track.

**Introduction.** First observed in 1985[1], the propagation of magnetic domain walls (DWs) by spin current has been a central topic in Spintronics. Early experiments used the *spin transfer torque* (STT) effect, through which a spin-polarized in-plane current drives a DW at large current densities (~100 GA/m$^2$), with large power consumption and Joule heating. A recent major advance in the efficiency of current-driven DWs was the use of perpendicularly-magnetized films with asymmetric interfaces adjacent to a heavy metal (HM) layer with strong spin-orbit coupling (SOC) (**Figure 1**a)[2,3]. Due to the spin Hall effect (SHE), a charge current in the HM layer induces a pure spin current that diffuses into the magnetic film, with density $J_S = \theta_{SHE} \frac{\hbar}{2e} J_C$ (where $\theta_{SHE}$ is the SHE angle and $J_C$ the charge current density), and with polarization direction **σ** perpendicular to both the **J**$_C$ and the film normal (in **Figure 1**a, **J**$_C \parallel$ **x** and **σ** = **y**). This spin current induces a torque, known as *damping-like spin-orbit torque* (DL SOT), that can be expressed as an equivalent field $\mathbf{H}_{DL} = \frac{\hbar J_S}{2eM_S t} \mathbf{m} \times \boldsymbol{\sigma}$, where **m** is the magnetization unit vector, $M_S$ the spontaneous magnetization, and $\hbar$ the reduced Planck constant.

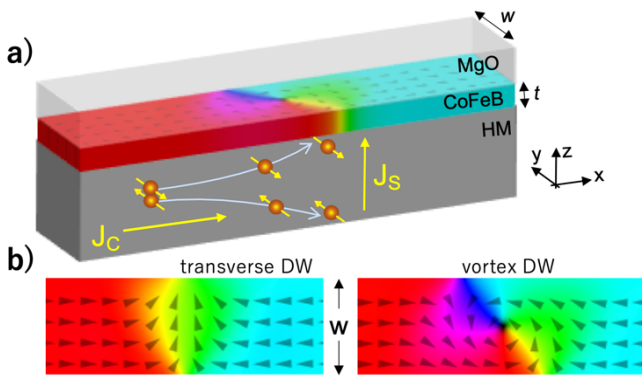

**Figure 1. a)** Schematic of SOT in a track of HM/CoFeB/MgO. A charge current $J_C$ flows mainly in the HM layer and induces a spin accumulation in the CoFeB layer. **b)** DW structures in in-plane magnetized soft tracks.

The DW propagation by DL SOT was discovered in perpendicularly-magnetized films and, so far, most studies have focused on this type of materials[4]. There is not, however, any fundamental reason precluding SOT from having an effect in in-plane-magnetized systems. Here, we use micromagnetic simulations and an analytical and numerical Thiele model to analyze the effects of SOT on DWs in in-plane-magnetized tracks, for different DW structures. Often, the $\mathbf{H}_{DL}$ at the center of the spin structure is used to evaluate the effect of SOT: it drives Néel DWs ($\mathbf{H}_{DL} \parallel \mathbf{z}$) and not Bloch DWs ($\mathbf{H}_{DL} = 0$). However, we show that this notion is misleading and that the effects of SOT depend fundamentally on the DW structure, in particular its cycloidal character.

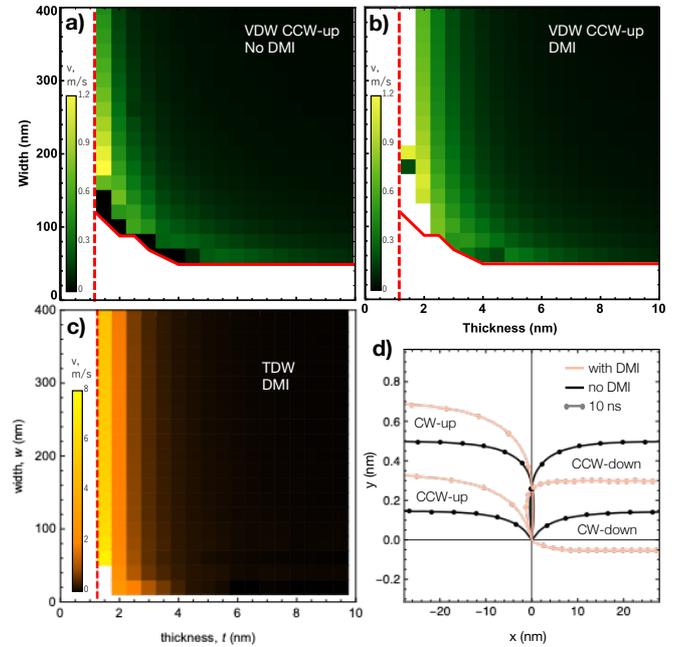

**Figure 2.** Micromagnetic simulations of SOT-driven DWs. **a,b,c)** Steady-state velocity versus $t,w$ ($J$ = 10 GA/m$^2$) for a CCW-up VDW with no DMI (a) and with DMI (b), and for a TDW with DMI (c). The red lines are the boundaries of the bi-stable region. Below the continuous red line, only TDW are stable. Left of the dashed red line, the domains are perpendicular. **d)** Trajectories of VDW cores with (black) and without (brown) DMI ($J$ = 50 GA/m$^2$, $t$ = 5 nm, $w$ = 150 nm). The points mark 10 ns intervals.

**DW structures in soft magnetic tracks.** A in-plane DW can be stabilized in one of two states, transverse domain wall (TDW) or vortex domain wall (VDW) (**Figure 1**b), depending on the chosen material and structure, especially on thickness ($t$) and the width ($w$) of the track. Note that VDWs can have clockwise or counter-clockwise winding (CW, CCW) and up (+z) or down (-z) core polarity, generally degenerate in energy. In this article, the labels CW-up or CW-down refer to VDW with CW winding and core polarity +z or -z, and similarly for CCW structures. Unless noted, DWs are head-to-head. The phase diagram for the magnetic configuration of DWs in in-plane magnetic nanotracks changing $t$ and $w$ in films with no anisotropy is known[5]. We determined this diagram for films with perpendicular anisotropy, with micromagnetic simulations using MuMax3[6]. We take the parameters of Pt/CoFeB/MgO[7–9]: $\mu_0 M_s = 1.7$ T, exchange stiffness $A_{ex} = 20$ pJ/m, and surface magnetic anisotropy $K_i = 1.3$ mJ/m². Starting from a TDW or a VDW (CW-down) on a track with given $t$ and $w$, the energy was minimized. We observed that for large $t$ and $w$ (up to 10 nm and 400 nm, respectively) both DW structures are (meta-)stable. However, for thin and narrow tracks, the VDW was not a stable state, meaning that TDW is exclusively stable in that region. The solid red line in **Figure 2**a separates the bistable and monostable regions. Below the spin reorientation thickness (1.2 nm), the magnetization is perpendicular (vertical red line in **Figure 2**a).

**SOT-driven dynamics.** Next, we study the dynamics of TDWs and VDWs under DL SOT in a track with $t = 5$ nm and $w = 150$ nm (damping constant $\alpha = 0.015$ [10], $\theta_{SHE} = 0.1$ [11], and no DMI). First, we observed that the TDW did not move, whereas the VDW moved with its core describing a trajectory that depended on its structure, shown in **Figure 2**d ($J = 50$ GA/m²; black lines). After a transient motion, the cores move steadily in +**x** (v= 0.5 m/s), which means that the VDWs can be driven along the track by DL SOT. VDWs with +**z** core polarity move leftwards and VDWs with -**z** move rightwards. During the initial transient motion, the VDW core moves transversally (in y) and, in some cases, backwards, until the repulsive force from track edge[12] limits the transversal deviation and leads to a steady state motion. This y deviation is larger for CW VDWs than for CCW, and is symmetric between up and down VDWs. We also observe that, if too much current is applied ($J \gtrsim 100$ GA/m²), the VDW core is expelled through the edge of the track and the VDW transforms into a TDW and stops propagating. Higher current leads to faster propagation in **x**, but also to a larger y-deviation. Therefore, it is possible to drive the VDWs with lower y deviation (CCW-up & CW-down) with higher current and velocity than the VDWs with higher y deviation.

Some HM layers, such as Pt, induce a significant DMI. We repeated the simulation of the core trajectory taking into account an interfacial DMI[13] of 0.9 pJ/m (brown lines in **Figure 2**d). We observe an increase in speed of the leftward-propagating VDWs along with a higher y deviation, and the opposite effect for rightward-propagating VDWs. In thinner tracks ($t \lesssim 2$ nm), which have a stronger effective DMI, the propagation direction of the down VDWs reverses and all VDWs move leftwards, although at different velocities. TDWs, which do not move without DMI, propagate leftwards with DMI ($v = 0.15$ m/s for $J = 10$ GA/m² and the same track dimensions as above).

We determined the ideal $w$ and $t$ to get the fastest DW propagation by repeating the simulation and calculating the final velocity for various $t$ and $w$, shown in **Figure 2**a-c for VDWs with and without DMI, and for TDW with DMI ($J = 10$ GA/m²). We have found that VDWs move faster for track dimensions near the borders between bistable and monostable regions and between in-plane and perpendicular anisotropy, up to 1.2 m/s ($t = 1.5$ nm, $w = 180$ nm). With DMI, the up VDWs move faster but the top velocity is still approximately the same (now at $t = 2$ nm, $w = 160$ nm), due to limitation caused by the expulsion of the vortex core at the edge, as discussed below. TDWs with DMI move faster in thinner tracks ($v = 0.15$ m/s for t = 5 nm and 5.8 m/s for $t = 2$ nm, $w = 150$ nm), but the TDW velocity varies little with track width, except in very narrow tracks ($w < 50$ nm) where they are faster (up to 8 m/s).

The DW mobility is lower than in perpendicular systems. For example, a Néel DW in a perpendicularly-magnetized track with the same parameters as **Figure 2**a except for $t=1$ nm, moves 38 times faster (46 m/s versus 1.2 m/s). We calculate, however, that the SOT-driven VDW is about five times faster than the same VDW driven by STT in a single CoFeB film (with current polarization $P = 0.5$ and non-adiabatic parameter $\beta = \alpha$).

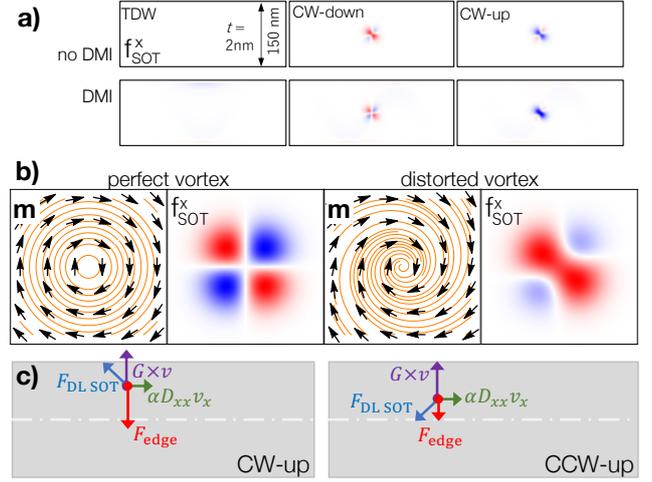

**Figure 3. Thiele forces on DWs. a)** $f^x_{SOT}$ for DWs with and without DMI (red is +x and blue is -x). **b)** Calculated $f^x_{SOT}$ for a perfect vortex (left) and a inwards-distorted vortex (right), with **m** shown in the stream plots. **c)** Schematic of Thiele forces for CCW and CW VDWs in the steady state.

**Trajectory analysis**. To study the DW trajectory analytically, we use the Thiele equation[14], which is derived from the Landau Lifshitz Gilbert equation[15] by assuming that the spin texture is rigid during motion:

$$\mathbf{G} \times \mathbf{v} - \alpha \mathbf{D} \mathbf{v} + \mathbf{F} = \mathbf{0} \quad (1)$$

where **G** is the gyrotropic vector, **v** the velocity, **D** the dissipation matrix, and **F** the sum of other forces like SOT ($\mathbf{F}_{SOT}$) or the repulsion of a VDW core by the track edge ($\mathbf{F}_{edge}$). These are all functions of a given spin distribution (see suppl. mat.). **G** only has a z component, $G_z = \pm 2\pi \frac{\mu_0 M_s}{\gamma_0} t$ for VDWs (sign given by the core polarization) or $G_z = 0$ for TDWs. **D** and $\mathbf{F}_{SOT}$ were calculated from the spin distributions at rest obtained with micromagnetic simulations, with $\mathbf{F}_{SOT}$ given by an integral over the track surface:

$$F^i_{SOT} = \int_S f^i_{SOT}\, dS = \int_S -\frac{\hbar j \theta_{SHE}}{2e}(\mathbf{m} \times \hat{\mathbf{y}}) \cdot \frac{\partial \mathbf{m}}{\partial i}\, dS \quad (2)$$

Where $e$ is the elementary charge and $i$ is $x$ or $y$. The repulsion of a VDW core by the edge is approximated by a Hooke's spring force[12] $\mathbf{F}_{edge} = -ky\hat{\mathbf{y}}$, where y is the core distance from the center of the track, and the spring constant $k$ is calculated from the balance of all other Thiele forces in a prior simulation of the DW motion (suppl. mat.). In that same simulation, we observed that the values of $G_z$, $D_{ij}$, and $F^i_{SOT}$ do not vary significantly during movement (<1%), which validates the use of the Thiele equation with constant parameters determined at rest. We then calculated the trajectories of a CW-down and a CCW-down VDWs with no DMI, which match very closely those from the micromagnetic simulations (suppl. mat.). The transient backwards motion that can be seen in some trajectories (Fig. 2d) can be understood as a gyrotropic effect associated with a significant $F_{SOT}^y$ (suppl. mat.).

In the steady state, the VDW propagates constantly along **x** with a velocity given by $v_x = \frac{F_{SOT}^x}{\alpha D_{xx}}$ under the Thiele forces sketched in **Figure 3**c. For CCW-up, the gyrotropic and $F_{SOT}^y$ forces are opposed, while for the CW-up these forces add up, which explains why the CW-up travels closer to the edge. $F_{SOT}^x$ is the origin of the steady state motion. $F_{SOT}^x = 0$ for a TDW with no DMI (Fig. 3a), and so the TDW does not propagate. This can be understood by noting that $\mathbf{m} \times \mathbf{y}$ is parallel to **z** while $\frac{\partial \mathbf{m}}{\partial x}$ is in-plane, so $f_{SOT}^x \propto -(\mathbf{m} \times \mathbf{y}) \cdot \frac{\partial \mathbf{m}}{\partial x} = 0$ (eq. 2). In contrast, $f_{SOT}^x$ for a VDW is non-zero near the vortex core (Fig. 3a). For comparison, $f_{SOT}^x$ is shown for a perfect vortex (Fig. 3b, left). While the $f_{SOT}^x$ for the perfect vortex is also non-zero, it totals $F_{SOT}^x = 0$ and there is no driving force. For the VDW, there are more positive components than negative and $F_{SOT}^x > 0$. The pattern of $f_{SOT}^x$ of the VDW is the same as the one for a vortex with an inwards distortion (Fig. 3b, right). This means that the efficiency of VDW driving depends on the degree of core distortion due to geometric confinement.

DMI significantly changes $F_{SOT}^x$, but does not affect any other Thiele parameters. While the $F_{SOT}^x$ due to geometric confinement changes sign between up and down or CW and CCW VDWs, adding DMI makes $F_{SOT}^x$ more negative for all DW types. This can be seen in the $f_{SOT}^x$ maps in Figure 3a, where DMI enhances the negative (blue) regions in all DW types. The effect of DMI can be understood as favoring a cycloidal spin rotation of fixed chirality, which tilts the central magnetization of the TDW and further distorts the VDW core, and directly affects the $F_{SOT}^x$ (see eq. 2). DMI then renders the up VDWs faster and the down VDWs slower, and the TDW mobile. At $t \lesssim 2$ nm, the DMI contribution is larger than the geometric distortion, changing the sign of $F_{SOT}^x$ for down VDWs. In this case, both up and down VDWs move leftwards, reversing the behavior without DMI shown in Figure 2a. For thicker tracks, the geometric contribution dominates, and up and down VDWs move in opposite directions.

The Thiele equation can be used to understand the variation of the velocity with track size (Fig. 2a,b). At fixed w, one could expect that $v_x = \frac{F_{SOT}^x}{\alpha D_{xx}} \propto 1/t$, as $D_{xx} \propto t$ (see suppl. mat.) and $F_{SOT}^x$ should be constant in $t$ without DMI (eq. 2) or even decrease with $t$ with DMI. The Thiele parameters calculated from simulations confirm the expected behavior (see suppl. mat.), except near the spin reorientation thickness, where $F_{SOT}^x$ strongly decreases with $t$ due to a decrease of the vortex core size. In wide tracks, $v_x$ decreases slightly with $w$, as $F_{SOT}^x$ is constant (being non-zero only in the core) and $D_{xx}$ increases slightly with $w$ (as it is most significant in the core). The variation of $v_x$ is much larger in narrow tracks, as the increasingly distorted core increases $F_{SOT}^x$ dramatically. So far, we treated head-to-head VDWs with -z core polarity. For the opposite core polarity, with no DMI, $G_z$, $F_{SOT}^x$, and $F_{SOT}^y$ change sign, and the VDW propagate leftwards. For tail-to-tail VDWs with no DMI, $G_z$ and $F_{SOT}^y$ remain the same, but $F_{SOT}^x$ changes sign. DMI contributes with a more negative $F_{SOT}^x$ in all cases.

**Velocity limitation of VDWs.** The Thiele equation can be used to estimate the maximum velocity limit imposed by the VDW expulsion. In the steady state, the final position in **y** should be less than the edge ($w/2$): $y_\text{final} = \frac{G_z v_x \pm F_{SOT}^y}{k} < \frac{w}{2}$, where the signs $+/-$ correspond to CCW/CW windings. The maximum current density and maximum velocity can be deduced:

$$J_{\max} = \frac{2 w k}{\frac{G_z}{\alpha D_{xx}} I_x \pm I_y}, \quad v_{\max} = \frac{w k}{2} \frac{1}{G_z \pm \alpha D_{xx} I_y / I_x} \quad (3)$$

where $I_i = |F_{SOT}^i|/J$. DMI only changes significantly $I_x$. Therefore, as in the micromagnetic simulations, we find that CW-down VDWs can be faster than CCW-down, as a larger current density can be applied without expelling the VDW. On the other hand, for up VDWs, it is the CCW-up that is the favorable configuration, as $F_{SOT}^x$ and $F_{SOT}^y$ have opposite sign and $G_z$ is negative. For tail-to-tail VDWs, $G_z$ and $F_{SOT}^y$ are the same as for head-to-head, but $F_{SOT}^x$ is of opposite sign. Therefore, e.g. a tail-to-tail CW-down VDW is unfavorable. While without DMI there are two optimal configurations, with DMI, one of them will be preferable. As DMI decreases $I_x$ for all DWs, the CCW-up moves the most efficiently. This velocity limitation can be overcome by using current pulses shorter than the expulsion time[16]. We observed in micromagnetic simulations that pulse-driven VDWs travel as fast as the steady state continuous current case and can then be driven at higher currents.

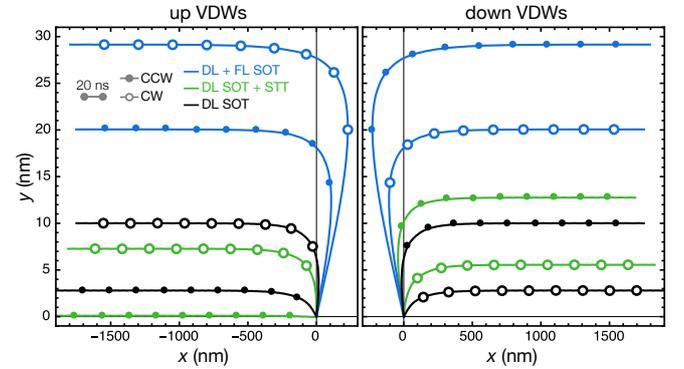

**Figure 4.** VDW core trajectory with DL SOT (black), DL & FL SOT (blue) and DL SOT & STT (green) for +z (left panel) or –z (right panel) core polarity. For all cases, $J = 1$ TA/m$^2$, $t = 5$ nm, $w = 150$ nm. Filled and open points correspond to CCW and CW winding, respectively. The points are separated by 20 ns.

**Effects of FL and STT**. As the FL SOT is mathematically equivalent to an external field along **y**, it cannot drive the DW in a steady state motion, but it can distort it and induce transient and reversible motion[17]. The ratio of magnitude of FL and DL SOT, $\xi$, depends on the HM (we use $\xi = 0.5$ for Pt[11]). Fig. 4 shows the trajectories of VDWs obtained from micromagnetic simulations under both DL and FL SOT (blue lines) and under DL SOT only (black), for comparison. As before, let us consider CW-down and CCW-down VDWs (right panel). We clearly see that, with FL SOT, both VDWs are pushed further up, and draw a curved path which protrudes further left. The FL SOT, being equivalent to an external field along +**y**, enlarges the region of the VDW with spins along +**y**, pushing the core of a head-to-head VDW upwards. The larger leftwards motion results from the gyrotropic effect and a larger initial $v_y$. After the transient motion, the steady state velocity is unchanged ($\lesssim 1\%$). The increase of y deviation reduces, however, the current threshold for VDW expulsion, and so a lower $\xi$ is preferable (i.e. Pt rather than Ta). For core polarity +**z**, which move leftwards, the same observations apply (left panel).

We study the effects of STT below the Walker threshold assuming, for simplicity, that the non-adiabatic STT is zero ($\beta = 0$) in the micromagnetic simulations. Considering Pt as the HM, the current density in CoFeB, $J_{FM}$, is estimated to be 10% of that in the Pt [18 Suppl. Mat.]. As before, we compare the trajectories under DL SOT and STT (green lines in Fig. 4) with those under DL SOT alone (in black), for the same four types of VDW. STT has a large effect on the motion in **y** direction. For –**z** core polarity, STT pushes the VDW farther up, decreasing the maximum current. On the other hand, STT restrains the y deviation for +z core polarity, increasing the maximum current. This difference can be explained with the Thiele equation adapted to include STT: $\boldsymbol{G} \times (\boldsymbol{v} - \boldsymbol{u}) - D(\alpha \boldsymbol{v} - \beta \boldsymbol{u}) + \boldsymbol{F} = \boldsymbol{0}$, where $\mathbf{u} = -\frac{\gamma \hbar P J_{FM}}{2 e M_s} \mathbf{e}_x$ is the STT drift velocity. As before, the $y$ position of the steady state can be deduced: $y_{final} = \left(G_z(v_x - u) \pm F_{SOT}^y\right)/k$. Therefore, STT either enhances or reduces the gyrotropic force in $y$ depending on the relative signs of $v_x$ and $u$. For -z core polarity, $v_x > 0$ and STT increases $y_{final}$, and vice-versa for +z core polarity. The effect of the winding, which determines the sign of $F_{SOT}^y$, is as before. Therefore, there is an optimal VDW configuration with minimal $y_{final}$ (CCW-up for our parameters).

The final velocity is $v_x = \frac{F_{SOT}^x}{\alpha D_{xx}} + \frac{\beta}{\alpha} u$. STT increases the final velocity of up VDWs and slows the down VDWs. Thus, it is possible to choose a configuration of VDW that minimizes the y deviation and maximizes velocity. An interesting consequence is that, by comparing the velocity of different VDW types in the same track (e.g. +z versus –z core polarity), it is possible to evaluate the relative magnitudes of SOT and STT in the material, for which no direct measurement exists.

**Conclusion.** DWs in in-plane magnetized tracks can be driven by SOT with an efficiency that depends on their structure, the size of the track, and the magnitude of DMI. The driving force arises from a cycloidal distortion of the DW structure caused by DMI or, in VDWs, also by geometrical confinement in thinner and narrower tracks. In the absence of DMI, only VDWs in thin and narrow tracks propagate. These findings underline that the effects of SOT may not be assessed simply from the $\mathbf{H}_{DL}$ at the center of the structure. The SOT drives different VDW types in opposite directions, which can be used to estimate the relative magnitude of SOT and STT from the mobilities of different VDWs types. Due to the gyrotropic force, the vortex core in VDW is pushed against the edge, which limits the maximum applicable current density. FL SOT increases the transversal deviation and thus limits further the VDW velocity. Using the parameters of the MgO/CoFeB/Pt system, we estimated that the VDW velocity by SOT may reach 1.2 m/s at 10 GA/m$^2$ for VDWs with or without DMI. This is three times faster than STT driven velocity, however 65 times smaller than the same system in the perpendicular magnetized configuration. For some in-plane DW structures, the effects of STT and SOT add up, resulting in an even higher velocity. These findings indicate that applications that require in-plane DWs, such as the manipulation of magnetic beads with DWs[19,20], may greatly benefit from combining SOT and STT.

**Acknowledgements.** We wish to thank R. Ferreira, D. Chiba, T. Koyama and R. Weil for preparing samples to obtain an experimental prospective and J. Miltat for useful discussions.

[1] P.P. Freitas and L. Berger, J. Appl. Phys. **57**, 1266 (1985).
[2] I.M. Miron, G. Gaudin, S. Auffret, B. Rodmacq, A. Schuhl, S. Pizzini, J. Vogel, and P. Gambardella, Nat. Mater. **9**, 230 (2010).
[3] A. Thiaville, S. Rohart, E. Jué, V. Cros, and A. Fert, Europhys. Lett. **100**, 57002 (2012).
[4] A. V. Khvalkovskiy, V. Cros, D. Apalkov, V. Nikitin, M. Krounbi, K.A. Zvezdin, A. Anane, J. Grollier, and A. Fert, Phys. Rev. B **87**, 020402 (2013).
[5] Y. Nakatani, A. Thiaville, and J. Miltat, J. Magn. Magn. Mater. **290–291**, 750 (2005).
[6] A. Vansteenkiste, J. Leliaert, M. Dvornik, M. Helsen, F. Garcia-Sanchez, and B. Van Waeyenberge, AIP Adv. **4**, 107133 (2014).
[7] M. Yamanouchi, A. Jander, P. Dhagat, S. Ikeda, F. Matsukura, and H. Ohno, IEEE Magn. Lett. **2**, 3000304 (2011).
[8] T. Devolder, J.-V. Kim, L. Nistor, R. Sousa, B. Rodmacq, and B. Diény, J. Appl. Phys. **120**, 183902 (2016).
[9] S. Ikeda, K. Miura, H. Yamamoto, K. Mizunuma, H.D. Gan, M. Endo, S. Kanai, J. Hayakawa, F. Matsukura, and H. Ohno, Nat. Mater. **9**, 721 (2010).
[10] X. Liu, W. Zhang, M.J. Carter, and G. Xiao, J. Appl. Phys. **110**, 033910 (2011).
[11] S. Emori, U. Bauer, S.-M. Ahn, E. Martínez, and G.S.D. Beach, Nat. Mater. **12**, 611 (2013).
[12] J. He, Z. Li, and S. Zhang, Phys. Rev. B - Condens. Matter Mater. Phys. **73**, 1 (2006).
[13] S. Tacchi, R.E. Troncoso, M. Ahlberg, G. Gubbiotti, M. Madami, J. Åkerman, and P. Landeros, Phys. Rev. Lett. **118**, 147201 (2017).
[14] A. Thiele, Phys. Rev. Lett. **30**, 230 (1973).
[15] T.L. Gilbert, IEEE Trans. Magn. **40**, 3443 (2004).
[16] J. Sampaio, S. Lequeux, P.J. Metaxas, A. Chanthbouala, R. Matsumoto, K. Yakushiji, H. Kubota, A. Fukushima, S. Yuasa, K. Nishimura, Y. Nagamine, H. Maehara, K. Tsunekawa, V. Cros, and J. Grollier, Appl. Phys. Lett. **103**, 242415 (2013).
[17] J.-Y. Chauleau, R. Weil, A. Thiaville, and J. Miltat, Phys. Rev. B **82**, (2010).
[18] S. DuttaGupta, S. Fukami, C. Zhang, H. Sato, M. Yamanouchi, F. Matsukura, and H. Ohno, Nat. Phys. 1 (2015).
[19] E. Rapoport and G.S.D. Beach, Sci. Rep. **7**, 10139 (2017).
[20] M. Donolato, P. Vavassori, M. Gobbi, M. Deryabina, M.F. Hansen, V. Metlushko, B. Ilic, M. Cantoni, D. Petti, S. Brivio, and R. Bertacco, Adv. Mater. **22**, 2706 (2010).